\documentstyle[prc,aps,preprint,epsfig]{revtex}
\tightenlines
\begin{document}
\draft
\title{Sound modes in hot nuclear matter}
\author{V.M.Kolomietz$^{1,2)}$ and S.Shlomo$^{1)}$}
\address{$^{1)}$ Cyclotron Institute, Texas A\&M University, College Station, 
TX 77843, USA}
\address{$^{2)}$ Institute for Nuclear Research, 03028 Kiev, Ukraine}
\maketitle

\begin{abstract}
The propagation of the isoscalar and isovector sound modes in a hot nuclear
matter is considered. The approach is based on the collisional kinetic
theory and takes into account the temperature and memory effects. It is
shown that the sound velocity and the attenuation coefficient are
significantly influenced by the Fermi surface distortion (FSD). The
corresponding influence is much stronger for the isoscalar mode than for the
isovector one. The memory effects cause a non-monotonous behavior of the
attenuation coefficient as a function of the relaxation time leading to a
zero-to-first sound transition with increasing temperature. The mixing of
both the isoscalar and the isovector sound modes in an asymmetric nuclear
matter is evaluated. The condition for the bulk instability and the
instability growth rate in the presence of the memory effects is studied. It
is shown that both the FSD and the relaxation processes lead to a shift of
the maximum of the instability growth rate to the longer wave length region.
\end{abstract}

\bigskip


\bigskip

\bigskip

\bigskip

\vskip 1cm

\section{Introduction}

In the vicinity of the equilibrium state nuclear matter is stable with
respect to particle density and Fermi surface distortions and the excitation
of both the isoscalar and the isovector sound modes is possible. Propagation
of an isoscalar sound wave in nuclear matter depends crucially on the Landau
parameter $F_{0}$ in the quasiparticle interaction amplitude and on the
relaxation processes. For zero temperature, there is an underdamped
zero-sound mode at $F_{0}>0$ because the phase velocity of the sound wave
exceeds the speed of particles inside the Fermi sphere. A strong Landau
damping appears at $-1<F_{0}<0$ where a nonzero transfer of energy from the
wave to particles is possible \cite{lipi} and the wave transforms to an
overdamped mode. The picture of propagation of the zero-sound wave is
essentially more complicated in the case of a hot nuclear matter. Due to the
existence of the temperature tail of the equilibrium distribution function,
the phase conditions for the Landau damping are fulfilled here for positive
values of the $F_{0}$ and a possibility for a propagation of the sound wave
appears in the region $-1<F_{0}<0$ \cite{koladi}.

The zero sound is transformed to the first-sound mode in the limit of strong
interaction, $|F_{0}|\gg 1,$ \cite{Peth} or in the frequent collision regime
at high temperatures \cite{abkh}. It is necessary to stress that the sound
velocity $c$ is directly related to the nuclear matter incompressibility
coefficient $K$ for the first-sound mode only. In this case, one has $%
c\approx c_{1}=\sqrt{K/9m}$, where $c_{1}$ is the velocity of the first
sound. In general, the sound velocity $c$\ is a complicated function of \
both $K$ and the dimensionless collisional parameter $\omega \tau $, where $%
\tau $ is the relaxation time and $\omega $ is the eigenfrequency of the
sound mode. In the present work we obtain a simple analytical expression for
the sound velocity $c$ which provides a description for both the frequent-
and rare-collision limit as well as for the intermediate cases. Special
attention is paid to the propagation of the isovector sound and for the
zero- to first- sound transition for this mode. It is well known \cite{aigr}
that hydrodynamic approaches like the Goldhaber-Teller or Steinwedel-Jensen
models give a reasonable description of the nuclear isovector giant
resonances (IVGR). However it is not the case for the isoscalar giant
resonances (ISGR), where the Fermi surface distortion effects plays an
important role \cite{Bert,HoEc79,NiSi80,Kolo83} and the traditional
hydrodynamic model can not be applied. In our approach, this situation is
directly related to the peculiarities of the propagation of both isovector
and isoscalar sounds.

The interparticle collisions on the distorted Fermi surface lead to the
collisional damping of the sound wave. Two limiting regimes $\omega \tau
\rightarrow 0$ and $\omega \tau \rightarrow \infty $ provide the existence
of the non-damped first and zero sounds, respectively, \cite{lipi}. In what
follows we will use the collisional kinetic theory, taking into account the
memory effects on the collision integral \cite{lipi,BeBo83,ayik,KoPl92}. We
will discuss a special feature of the temperature dependence of the
attenuation of the sound mode in hot nuclear matter. In particular, we will
show that memory effects lead to a bell-shaped form of the attenuation
coefficient $\kappa $ as a function of the temperature, providing a correct
behavior of $\kappa $\ in both limiting regimes $\omega \tau \rightarrow 0$
and $\omega \tau \rightarrow \infty $.

With decreasing bulk density or increasing temperature the nuclear matter
reaches the regions of mechanical or thermodynamical instabilities with
respect to small particle density distortions \cite
{Peth,peth87,CoCh94,CoCh94a,AyCo95,BaCo98} and to separation into liquid and
gas phases \cite{JaMe83,MuSe95}. The general instability condition of the
Fermi liquid reads $1+F_{k}/(2k+1)<0$ \cite{Pome}, where $F_{k}$ is the
Landau's parameter in the expansion of the quasiparticle interaction
amplitude in Legendre polynomial \cite{lipi}. However the development of
instability depends not only on the equation of state, but also on the
dynamical effects such as the dynamical Fermi-surface distortion ({\rm FSD})
effect \cite{kosh99}. The FSD effects strongly reduce, by the factor $\sim
(|F_{0}|-1)^{1/2},$\ the instability growth rate $\Gamma $ in the unstable
region $0<-1-F_{0}\ll 1$ \cite{Peth}. In the present work, we will consider
the influence of the Fermi-surface distortion, relaxation and memory effects
on the instability growth rate $\Gamma $.

The plan of the paper is as follows. In Sec. II we derive the general
equation of motion for the particle density vibrations in the presence of
the memory effects. We introduce the renormalized incompressibility
coefficient and the viscosity coefficient which are frequency dependent due
to the memory effect. In Sec. III we obtain a simple expression for the
refraction and attenuation coefficients for both the isoscalar and the
isovector sound waves which propagate in a hot nuclear matter. We consider
also the mixing of the isoscalar and isovector modes in an asymmetric
nuclear matter taking into account the relaxation and the memory effects.
The development of the bulk instability in a low density nuclear matter in
the presence of the relaxation and the memory effects is considered in Sec.
IV. The conclusion is given in Sec. V.

\section{Sound propagation in hot nuclear matter}

We start from the collisional kinetic equation \cite{lipi} 
\begin{equation}
{\frac{\partial f}{\partial t}}+{\frac{{\bf p}}{m}}{\frac{\partial f}{%
\partial {\bf r}}}-{\frac{\partial V}{\partial {\bf r}}}{\frac{\partial f}{%
\partial {\bf p}}}={\rm \delta St}(t).  \label{2.1}
\end{equation}
\noindent Here ${\rm \delta St}(t)$\ is the collisional integral, $V\equiv V(%
{\bf r},t)$ is the self-consistent mean field and $f$ $\equiv f({\bf r,p},t)$
is the Wigner distribution function in which we will take into account only
the distortion of the Fermi sphere with multipolarities $\ell \leq 2$ 
\begin{equation}
f=f_{s}+\delta f,\ \ \ \delta f=\sum_{\ell =1}^{2}\delta f_{\ell }.
\label{2.2}
\end{equation}
\noindent The distribution $f_{s}\equiv f_{s}({\bf r,p},t)$ corresponds to
the spherical Fermi surface and $\delta f$ represents both the quadrupole
deformation and the displacement of the Fermi surface. For small deviations
from a Fermi sphere the right-hand side (RHS) ${\rm \delta St}(t)$ of (\ref
{2.1}) is a collision integral linearized in $\delta f$ and it can be
represented in the form 
\begin{equation}
{\rm \delta St}(t)=\int_{-\infty }^{t}dt^{\prime }A(t-t^{\prime })\delta
f(t^{\prime }),  \label{2.3}
\end{equation}
which takes into account the memory effects due to the memory kernel $%
A(t-t^{\prime })$. In this paper we will not use an explicit form of ${\it A}
$. Below we will need ${\rm \delta St}(t)$ only for periodic oscillation of $%
\delta f$ \ with the eigenfrequency $\omega .$ Assuming the restriction $%
\ell \leq 2$ for the Fermi surface distortion, the collision integral (\ref
{2.3}) can be written in the form of the extended $\tau $-approximation, see
Refs. \cite{koplsh1,KiKo96},

\begin{equation}
{\rm \delta St}(t)=-\frac{\delta f_{2}}{\tau _{r,\omega }},  \label{2.tau1}
\end{equation}
where the relaxation time $\tau _{r,\omega }$ is $\omega $-dependent\ due to
the memory effects.

The assumption in Eq. (\ref{2.2}) allows us to reduce the collisional
kinetic equation (\ref{2.1}) to the local equations of motion for the
particle density $\rho \equiv \rho ({\bf r},t),$ the displacement field $%
\chi _{\alpha }\equiv \chi _{\alpha }({\bf r},t)$ and the pressure tensor $%
P_{\alpha \beta }\equiv P_{\alpha \beta }({\bf r},t)$ derived as the lowest $%
{\bf p}$-moments of the distribution function $f({\bf r,p},t)$:

\begin{equation}
\rho =\int {\frac{g\ d{\bf p}}{(2\pi \hbar )^{3}}}f,\quad \frac{\partial
\chi _{\alpha }}{\partial t}={\frac{1}{\rho }}\int {\frac{g\ d{\bf p}}{(2\pi
\hbar )^{3}}}{\frac{p_{\alpha }}{m}}f,\quad P_{\alpha \beta }={\frac{1}{m}}%
\int {\frac{g\ d{\bf p}}{(2\pi \hbar )^{3}}}(p_{\alpha }-mu_{\alpha
})(p_{\beta }-mu_{\beta })f.  \label{2.4}
\end{equation}
Here ${u}_{\alpha }=\partial {\chi }_{\alpha }/\partial t$ is the velocity
field and $g$ is the spin-isospin degeneracy factor. Taking the first two $%
{\bf p}$-moments of Eq. (\ref{2.1}) one obtains, see Refs. \cite
{koplsh1,KiKo96,kota},

\begin{equation}
m\rho {\frac{\partial ^{2}}{\partial t^{2}}}\chi _{\nu }+({\frac{\partial }{%
\partial r_{\nu }}}P+\rho {\frac{\partial }{\partial r_{\nu }}}V)+{\frac{%
\partial }{\partial r_{\mu }}}P_{\nu \mu }^{\prime }=0,  \label{2.5}
\end{equation}
where $P$ is the pressure due to motion of nucleons without distortion of
the Fermi sphere and $P_{\nu \mu }^{\prime }$ is associated with quadrupole
distortion of the Fermi surface: $P_{\alpha \beta }=P\delta _{\alpha \beta
}+P_{\alpha \beta }^{\prime }.$ The pressure tensor $P_{\nu \mu }^{\prime }$
is responsible for the dissipative processes. Taking the second ${\bf p}$%
-moment of Eq. (\ref{2.1}) one obtains the following equation for the
pressure tensor

\begin{equation}
{\frac{\partial }{\partial t}}P_{\alpha \beta }+{\frac{\partial }{\partial r}%
}_{\nu }u_{\nu }P_{\alpha \beta }+P_{\nu \beta }{\frac{\partial }{\partial r}%
}_{\nu }u_{\alpha }+P_{\nu \alpha }{\frac{\partial }{\partial r}}_{\nu
}u_{\beta }=-P_{\alpha \beta }^{\prime }/\tau _{r,\omega }.  \label{2.6}
\end{equation}

The equations of motion (\ref{2.5}) and (\ref{2.6}) are closed. They can be
applied to both the isoscalar and isovector sound excitations. Let us
consider the isoscalar compression mode. To simplify the problem, we can
rewrite the expression in the parenthesis in Eq. (\ref{2.5}) near the
equilibrium value of the density $\rho _{{\rm eq}}$\ as 
\begin{equation}
{\frac{\partial }{\partial r_{\nu }}}P+\rho {\frac{\partial }{\partial
r_{\nu }}}V=\rho {\frac{\partial }{\partial r_{\nu }}}{\frac{\delta \epsilon 
}{\delta \rho }}\approx \rho _{{\rm eq}}{\frac{\partial }{\partial r_{\nu }}}%
\left[ \left( {{\frac{\delta ^{2}\epsilon }{\delta \rho ^{2}}}}\right) _{%
{\rm eq}}\delta \rho \right] ,  \label{2.7}
\end{equation}
where index ''${\rm eq}$'' refers the equilibrium state, $\epsilon $ is the
energy density of particles 
\begin{equation}
\epsilon =\epsilon _{{\rm kin}}+\epsilon _{{\rm pot}},  \label{2.8}
\end{equation}
$\epsilon _{{\rm kin\ }}$ is the kinetic energy density

\begin{equation}
\epsilon _{{\rm kin}}=\frac{3}{2}P=\int {\frac{g\ d{\bf p}}{(2\pi \hbar )^{3}%
}}\,{\frac{p^{2}}{2m}}f_{s}={\frac{3}{10}}{\frac{\hbar ^{2}}{m}}{({\frac{3{%
\pi }^{2}}{g}})}^{2/3}\rho ^{5/3}  \label{2.9}
\end{equation}
and\ $\epsilon _{{\rm pot}}$ is the potential energy density which is
related to the mean field $V$ by 
\begin{equation}
V=\delta \epsilon _{{\rm pot}}/\delta \rho .  \label{2.10}
\end{equation}
Note that $(\delta \epsilon /\delta \rho )_{{\rm eq}}$ is the chemical
potential, which does not depend on the space coordinate ${\bf r,}$ for the
equilibrium state of the nucleus. We have used this fact when deducing Eq. (%
\ref{2.7}). Solving Eq. (\ref{2.6}) with respect to $P_{\alpha \beta
}^{\prime },$ using Eq. (\ref{2.7}) and the continuity equation $\delta \rho
=\rho -\rho _{{\rm eq}}=-{\rm div}(\rho _{{\rm eq}}{\bf \chi })$, we obtain
an equation for the density vibration in the form

\begin{equation}
\omega ^{2}\delta \rho +(K_{\omega }^{\prime }/9m)\nabla ^{2}\delta \rho
=i\omega (4\eta _{\omega }/3m\rho _{{\rm eq}})\nabla ^{2}\delta \rho .
\label{2.11}
\end{equation}
Here

\begin{equation}
K_{\omega }^{\prime }=K+8{(\epsilon _{kin}/\rho )}_{{\rm eq}}{\rm 
\mathop{\rm Im}%
}\left( \frac{\omega \tau _{r,\omega }}{1-i\omega \tau _{r,\omega }}\right) ,
\label{2.13}
\end{equation}
where $K\equiv 9{(\delta ^{2}\epsilon /\delta \rho ^{2})}_{{\rm eq}}\rho _{%
{\rm eq}}$ is the static incompressibility and $\eta _{\omega }$ is the
viscosity coefficient

\begin{equation}
\eta _{\omega }=%
\mathop{\rm Re}%
\left( \frac{\tau _{r,\omega }}{1-i\omega \tau _{r,\omega }}\right) P_{{\rm %
eq}}.  \label{2.14}
\end{equation}
We point out \ that there is a significant difference between the static
nuclear incompressibility coefficient, $K$, i.e., derived as a stiffness
coefficient with respect to a change in the bulk density, and the dynamic
one, $K_{\omega }^{\prime }$ of Eq. (\ref{2.13})$,$ associated with the
sound propagation. This difference is due to the second term on the RHS of
Eq. (\ref{2.13}) caused by the Fermi-surface distortion effects. The
quantity $\eta _{\omega }$ in Eq. (\ref{2.14}) determines the time
irreversible contribution to the pressure tensor $P_{\alpha \beta }^{\prime }
$ and can be considered as the viscosity coefficient due to the relaxation
occurring on the distorted Fermi surface. Expression (\ref{2.14}) is valid
independently of the nucleon's collision rate. The viscosity goes to zero in
both the rare, $\tau _{r,\omega }\rightarrow \infty ,$ and frequent, $\tau
_{r,\omega }\rightarrow 0,$ collision limits.

\section{Dispersion relation, memory effects and damping}

Assuming a plane wave solution $\delta \rho \sim \exp (i{\bf q\cdot r-}%
i\omega t)$ one obtains from Eq. (\ref{2.11}) the following dispersion
relation

\begin{equation}
\omega ^{2}=(K_{\omega }^{\prime }/9m)q^{2}-i\omega (4\eta _{\omega }/3m\rho
_{{\rm eq}})q^{2}.  \label{3.1}
\end{equation}
The solution of this equation defines the complex wave number $q$\ ($\omega $
is real). A simple solution to Eq. (\ref{3.1}) can be obtained in two
limiting cases of the frequent collision (first sound) regime, $\omega \tau
_{r,\omega }\rightarrow 0,$ and the rare collision (zero sound) regime, $%
\omega \tau _{r,\omega }\rightarrow \infty .$ The sound velocity $c=\omega
/q $ is given by

\begin{equation}
c=c_{1}=\sqrt{K/9m}\ \ {\rm if}\ \ \omega \tau _{r,\omega }\rightarrow
0\quad {\rm and}\quad c=c_{0}=\sqrt{(K+\Delta K)/9m}\quad {\rm if}\quad
\omega \tau _{r,\omega }\rightarrow \infty ,  \label{3.cc}
\end{equation}
where $\Delta K\approx 8{(\epsilon _{{\rm kin}}/\rho )}_{{\rm eq}}\approx
(24/5)\ e_{F}\approx 200\ {\rm MeV}$ (we adopted the kinetic Fermi energy $%
e_{F}\approx 40\ {\rm MeV}$). We point out that the value of $\Delta K$ is
comparable with the static incompressibility $K\approx 220$ {\rm MeV }$\ $%
and we have $c_{0}\approx \sqrt{2}c_{1}$. The factor $\sqrt{2}$ in this
relation is due to the restriction $\ell \leq 2$ for the multipolarity $\ell 
$ of the Fermi surface distortion. In a general case of arbitrary $\ell $
this factor is increased to $\sqrt{3}$ \cite{lipi}. The result (\ref{3.cc})
means that in contrast to the first sound (frequent collision) regime, the
sound velocity of the compression mode can not, in general, be used directly
to extract the static incompressibility of $K$ because of the additional
contribution from the Fermi surface distortion effects which result in the
renormalization of the incompressibility $K$ into $K_{\omega }^{\prime }$.

Using both asymptotic sound velocity $c_{1}$ and $c_{0},$ the solution to
the dispersion relation (\ref{3.1}) can be written as 
\begin{equation}
q=\frac{\omega }{c_{0}}(n+i\kappa ),  \label{3.qq}
\end{equation}
where the refraction coefficient $n$ and the attenuation coefficient $\kappa 
$ (both real)\ are obtained from the following equation 
\begin{equation}
n+i\kappa =\sqrt{\frac{1-i\omega \tau _{r,\omega }}{(c_{1}/c_{0})^{2}-i%
\omega \tau _{r,\omega }}.}  \label{3.nkapa}
\end{equation}
In the frequent collision (first sound) regime we obtain from Eq. (\ref
{3.nkapa})

\begin{equation}
n=\frac{c_{0}}{c_{1}},\qquad \kappa =\omega \tau _{r,\omega
}(c_{0}/2c_{1})[(c_{0}/c_{1})^{2}-1]\qquad {\rm if\ \ \ \ \ }\omega \tau
_{r,\omega }\ll 1.  \label{3.nkapa1}
\end{equation}
In the opposite case of the rare collision (zero sound) regime we obtain 
\begin{equation}
n=1,\qquad \kappa =[1-(c_{1}/c_{0})^{2}]/(2\omega \tau _{r,\omega })\qquad 
{\rm if\ \ \ \ \ }\omega \tau _{r,\omega }\gg 1.  \label{3.nkapa0}
\end{equation}

The attenuation coefficient $\kappa $ in both limiting regimes is a
complicated function of the frequency $\omega $ because of the memory effect
in the relaxation time $\tau _{r,\omega }$. In the case of sound propagation
in hot nuclear matter, the competition between the temperature smoothing
effects in the equilibrium distribution function and dynamical distortions
of the particle momentum distribution leads to the following expression for
the relaxation time \cite{lipi,ayik,KoPl92}

\begin{equation}
\tau _{r,\omega }={\frac{\tau _{0}}{{T^{2}+\xi \,(\hbar \omega )^{2}}}}~,
\label{3.tau}
\end{equation}
where $T$ is the temperature of nuclear matter and the $\omega $-dependence
of $\tau _{r,\omega }$ is due to the memory effects in the collision
integral. Below we will use ${\xi }=1/4\pi ^{2}$ \thinspace \cite{lipi}
and\thinspace $\tau _{0}=\alpha \,\hbar ,\,\alpha =9.2\,{\rm MeV}$ (for
the isoscalar mode) \thinspace \cite{KoPl96}.

Equations (\ref{3.1}) and (\ref{3.nkapa}) are valid for arbitrary collision
times $\tau _{r,\omega }$ and thus describe both the zero and the first
sound limit as well as the intermediate cases. From it one can obtain the
leading order terms in the different limits mentioned. In Fig. 1 we have
plotted both coefficients $n$ and $\kappa $ as obtained from Eq. (\ref
{3.nkapa}). In the high temperature limit, the system goes to the frequent
collision (first sound) regime with the saturated refraction coefficient $n$ 
$\approx c_{0}/c_{1}\approx \sqrt{3}$ (we use the factor $\sqrt{3}$ instead
of $\sqrt{2}$ assuming the contribution of the higher multipolarities $\ell
>2$ \ \ in the Fermi surface distortion as was mentioned above) and the
attenuation coefficient $\kappa \sim \tau _{r,\omega }\sim 1/T^{2}$. In the
opposite low temperature limit, the system is close to the zero sound regime
with $n\approx 1.$ We point out a shift of both $n$ and $\kappa $ by nonzero
values at $T\rightarrow 0.$ This is due to the memory effect in the
relaxation time $\tau _{r,\omega }$ of Eq. (\ref{3.tau}): in the very high
frequency limit, the system can exist close to the first sound regime at $%
n\approx \sqrt{3}$ even at zero temperature. The position of the maximum of $%
\kappa (T)$ in Fig. 1 can be interpreted as the transition temperature $T_{%
{\rm tr}}$ of zero- to first- sound regimes in a hot Fermi system. The value
of $T_{{\rm tr}}$ depends slightly on the sound frequency $\omega $ and it
is shifted to smaller values with the increase of $\omega .$

Let us consider now the isovector sound mode in a symmetric nuclear matter
with $\rho _{n,{\rm eq}}=\rho _{p,{\rm eq}},$ where $\rho _{n,{\rm eq}}$ and 
$\rho _{p,{\rm eq}}$ are the equilibrium neutron and proton density
respectively. The general equations of motion (\ref{2.5}) and (\ref{2.6})
are still correct. However the energy density $\epsilon $\ in Eq. (\ref{2.7}%
) is related now to the symmetry energy $E_{{\rm symm}}$. The corresponding
first sound velocity $c_{1}$ for the isovector mode is given by, \cite{bomo2}
Ch. 6, 
\begin{equation}
c_{1}=\sqrt{2E_{{\rm symm}}/m},  \label{3.10}
\end{equation}
where $E_{{\rm symm}}=(1/3)e_{F}(1+F_{0}^{\prime })\approx 30$ MeV, and $%
F_{0}^{\prime }$\ is the isovector Landau parameter in the quasiparticle
interaction amplitude. The zero sound velocity $c_{0}$ for the isovector
mode can be found from Eqs. (\ref{2.5}) and (\ref{2.6}) in the rare
collision limit $\tau _{r,\omega }\rightarrow \infty .$ Taking into account
Eq. (\ref{2.10}), one obtains, see also Eq. (\ref{3.cc}),

\begin{equation}
c_{0}=\sqrt{2(E_{{\rm symm}}+\Delta E_{{\rm symm}})/m},  \label{3.11}
\end{equation}
where $\Delta E_{{\rm symm}}\approx (4/9){(\epsilon _{kin}/\rho )}_{{\rm eq}%
}\approx (4/15)\ e_{F}\approx 10\ {\rm MeV.}$ We point out that, in contrast
to the isoscalar mode, the Fermi surface distortion effect leads to a
relatively small increase of the isovector zero sound velocity $c_{0},$ Eq. (%
\ref{3.11}),\ with respect to the first sound one $c_{1},$ Eq. (\ref{3.10}).
The dispersion relation (\ref{3.1}) takes the form 
\begin{equation}
\omega ^{2}=(2E_{{\rm symm,}\omega }^{\prime }/m)q^{2}-i\omega (4\eta
_{\omega }/3m\rho _{{\rm eq}})q^{2},  \label{3.12}
\end{equation}
where 
\begin{equation}
E_{{\rm symm,}\omega }^{\prime }=E_{{\rm symm}}+(4/9){(\epsilon _{kin}/\rho )%
}_{{\rm eq}}{\rm Im}\left( \frac{\omega \tau _{r,\omega }}{1-i\omega \tau
_{r,\omega }}\right) .  \label{3.13}
\end{equation}
All relations (\ref{3.qq})-(\ref{3.tau}) are still correct in the isovector
case if both asymptotic velocity $c_{0}$\ and $c_{1}$\ are taken from Eqs. (%
\ref{3.10}) and (\ref{3.11}). In Fig. 2 we have plotted the coefficients $n$
and $\kappa $ as obtained from Eq. (\ref{3.nkapa}) for the isovector mode
with the collision parameter $\alpha =4.6$ {\rm MeV} \thinspace \cite{KoPl96}%
. We point out that the transition temperature $T_{{\rm tr}}$ of zero- to
first- sound regimes for the isovector mode is significantly smaller than $%
T_{{\rm tr}}$ for the isoscalar one.

In an asymmetric nuclear matter, both the isovector and the isoscalar modes
are dependent on each other. The particle density fluctuation $\delta \rho $
takes a bispinor form $\delta \rho =(\delta \rho _{+},$ $\delta \rho _{-})$,
where $\delta \rho _{+}$ and $\delta \rho _{-}$ are the isoscalar and
isovector components respectively. A solution of the corresponding equations
of motion (\ref{2.5}) and (\ref{2.6}) leads to the following dispersion
relation, see also Eqs. (\ref{3.1}) and (\ref{3.12}),

\begin{equation}
{\rm Det}\left( 
\begin{array}{cc}
\omega ^{2}-(K_{\omega }^{\prime }/9m)q^{2}+i\omega (4\eta _{\omega }/3m\rho
_{{\rm eq}})q^{2} & (Ix/m)q^{2}+{\cal O}(I^{2}) \\ 
(Iy/m)q^{2}+{\cal O}(I^{2}) & \omega ^{2}-(E_{{\rm symm,}\omega }^{\prime
}/9m)q^{2}+i\omega (4\eta _{\omega }/3m\rho _{{\rm eq}})q^{2}
\end{array}
\right) =0.  \label{3.14}
\end{equation}
Here, $I=(\rho _{n}-\rho _{p})_{{\rm eq}}/(\rho _{n}+\rho _{p})_{{\rm eq}%
}\ll 1$ is the asymmetry parameter, $\rho _{n}$ and $\rho _{p}$ are the
neutron and proton densities respectively, $\rho _{{\rm eq}}=(\rho _{n}+\rho
_{p})_{{\rm eq}}$ and the coupling constants $x$ and $y$ are given by

\begin{equation}
x=-(2/9)(4\epsilon _{F}-K/3),\qquad y=-(2/9)(4\epsilon _{F}+K/6-9E_{{\rm symm%
}}).  \label{3.15}
\end{equation}
As is seen from Eq. (\ref{3.15}), the eigenfrequency $\omega $\ and the
corresponding sound velocity for both the isoscalar and the isovector modes
are independent of each other in the linear order of the asymmetry parameter 
$I.$ The structure of bispinor $\delta \rho =(\delta \rho _{+},$ $\delta
\rho _{-})$ is different for both the isoscalar-like and the isovector-like
modes. For the isoscalar-like mode with the eigenfrequency $\omega _{{\rm is}%
}$ given by a solution to Eq. (\ref{3.1}), the main contribution to the
bispinor $\delta \rho $ is due to the isoscalar component $\delta \rho
_{+}\sim 1$ while the isovector component $\delta \rho _{-}$ is proportional
to the asymmetry parameter$\ I.$ Namely, 
\begin{equation}
\left( \frac{\delta \rho _{-}}{\delta \rho _{+}}\right) _{{\rm is}}=\frac{%
yq^{2}/m}{(\omega ^{2}-(E_{{\rm symm,}\omega }^{\prime }/9m)q^{2}+i\omega
(4\eta _{\omega }/3m\rho _{{\rm eq}})q^{2})_{\omega =\omega _{{\rm is}}}}I.
\label{3.16}
\end{equation}
The opposite situation takes place for the isovector-like mode with the
eigenfrequency $\omega _{{\rm iv}}$ given by a solution to Eq. (\ref{3.12}).
In this case, one has $\delta \rho _{-}\sim 1$ and $\delta \rho _{+}\sim I.$
\ 

\section{Bulk instability}

Let us consider now the bulk instability regime $K<0$ and introduce an
instability growth rate $\Gamma =-i\,\omega $ ($\Gamma $ is real, $\Gamma >0$%
), see Refs. \cite{peth87,kosh99}. The amplitude of the density
fluctuations, $\delta \rho \sim \exp (i{\bf q\cdot r-}i\omega t)\sim \exp
(\Gamma t),$ grows exponentially if $\Gamma >0$. To prevent an unphysical
infinite growth of the short wave length fluctuations (see dotted line in
Fig. 3), we will take into account the velocity dependent contribution to
the effective interparticle interaction. Due to the corresponding change in
the selfconsistent mean field $V$ in Eq. (\ref{2.5}), an additional
anomalous term $\sim q^{4}$ appears in the dispersion relation (\ref{3.1})
and the equation for the instability growth rate $\Gamma $ takes the
following form \cite{kosh99} 
\begin{equation}
\Gamma ^{2}=|c_{1}|^{2}\,q^{2}-\zeta (\Gamma )\,q^{2}-\kappa _{s}\,q^{4}.
\label{4.1}
\end{equation}
Here $c_{1}=i\sqrt{|K|/9m}$ and 
\begin{equation}
\zeta (\Gamma )={\frac{4}{3\,m}}{\frac{\Gamma \tau _{r}}{{1+\Gamma \tau }_{r}%
}e_{F},}  \label{4.2}
\end{equation}
where $\tau _{r}$ is the relaxation time $\tau _{r}=\alpha \hbar /T^{2}$ 
\cite{abkh}. The constant $\kappa _{s}$ in the anomalous dispersion term in
Eq. (\ref{4.2}) depends on the model. In the case of the effective Skyrme
forces, one has \cite{kosh99} $\kappa _{s}=\hbar
^{2}/9m^{2}+(9t_{1}-5t_{2})\rho _{0}/32m,$ where $t_{1}$ and $t_{2}$ are the
parameters of the velocity dependent part of the Skyrme forces \cite{vabr}.
We point out that the instability regime with $K<0$ can be reached at a low
bulk density $\rho _{0}.$ The incompressibility $K$ is given by 
\begin{equation}
K=6\,e_{F}\,(1+F_{0})\,\left( 1+F_{1}/3\right) ^{-1}.  \label{4.3}
\end{equation}
Here, the Landau parameters $F_{k}$ are related to the parameters $t_{n}$ of
the effective Skyrme forces. Namely, 
\begin{equation}
F_{0}={\frac{9\,\rho _{0}}{8\,e_{F}}}\,\left[ t_{0}+{\frac{3}{2}}%
\,t_{3}\,\rho _{0}\right] \,{\frac{m^{\ast }}{m}}+3\,\left( 1-{\frac{m^{\ast
}}{m}}\right) ,\,\,\,F_{1}=3\,\left( {\frac{m^{\ast }}{m}}-1\right) ,
\label{4.4}
\end{equation}
where ${m/}m^{\ast }=1+m\,\rho _{0}\,(3\,t_{1}+5\,t_{2})/8\,\hbar ^{2}.$ For
the commonly used set of parameters $t_{n}$, the instability regime with $%
F_{0}<-1$ is reached at $\rho _{0}\lesssim 0.5\rho _{{\rm eq}}.$

In Fig. 3 we have plotted the instability growth rate $\Gamma $ as obtained
from Eq. (\ref{4.1}). The calculation was performed for the Skyrme force
SIII. For the relaxation time $\tau _{r}$ we used $\alpha =9.2$ $\,$MeV
and the bulk density $\rho _{0}$ was $\rho _{0}=0.3\,\rho _{{\rm eq}}$,
where $\rho _{{\rm eq}}$ is the saturated density $\rho _{{\rm eq}}=0.1453$ $%
{\rm fm}^{-3}$. We also show in Fig. 3 the result for the nonviscous nuclear
matter neglecting the anomalous dispersion term $\sim q^{4}$ and the Fermi
surface distortion effect (dotted line). The non-monotonous behavior of the
instability growth rate as a function of the wave number $q$ is due to the
anomalous dispersion term in Eq. (\ref{4.1}), induced by the velocity
dependent terms in the interparticle interaction. The instability growth
rate $\Gamma $ reaches a maximum, $\Gamma _{\max },$ at a certain $q=q_{\max
}$ and $\Gamma $ goes to zero at $q=q_{{\rm crit}}.$ The existence of the
critical wave number $q_{{\rm crit}}=|c_{1}|^{2}/\kappa _{s}$ for an
unstable mode is a feature of the system with the anomalous dispersion term 
\cite{peth87}.\ The distortion of the Fermi surface leads to a decrease of
the critical value $q_{{\rm crit}}$, i.e., the nuclear matter becomes more
stable due to the FSD effect. We can also see that the presence of viscosity
and the FSD effect lead to a shift of the position $q_{\max }$\ of the
maximum of $\Gamma (q)$ to the left. Thus, the instability of the nuclear
matter with respect to short-wave-length density fluctuations decreases due
to the viscosity and the FSD effect and the most unstable mode is shifted to
the region of the creation of larger clusters in the disintegration of
nuclear matter. For a saturated nuclear liquid one has for the force
parameters $t_{0}<0,\,\,\,t_{3}>0$ and $t_{s}>0$. However both values $%
q_{\max }$ and $q_{{\rm crit}}$\ have a non-monotonous behavior as a
function of the bulk density $\rho _{0}$ because of the additional $\rho
_{0}-$dependence of the Fermi energy $e_{F}$ in Eqs. (\ref{4.3}) and (\ref
{4.4}). The particle density dependence of the values of $q_{\max }$ and $q_{%
{\rm crit}}$ is shown in Fig. 4. The instability growth rate $\Gamma (q)$\
as well as the values of $q_{\max }$ and $q_{{\rm crit}}$ are only slightly
sensitive to the change of temperature at $T\lesssim 10${\rm \ MeV, }where
the temperature dependence of the bulk density $\rho _{0}$ can be neglected.
A more sophisticated consideration is necessary near the critical
temperature $T_{{\rm crit}}\approx 17$ {\rm MeV} where the nuclear matter is
unstable with respect to the liquid-gas\ phase transition.

\section{Conclusion}

Starting from the collisional kinetic equation (\ref{2.1}), we have derived
the dispersion relations (\ref{3.14}) and (\ref{4.1}) for both the stable
and the unstable regime of the density fluctuations in a heated nuclear
matter. The dispersion relations are influenced strongly by the FSD effect
and the anomalous dispersion term. The presence of the Fermi surface
distortion enhances the stiffness coefficient for a stable mode and reduces
the instability growth rate for an unstable one. There is a significant
difference between the static nuclear incompressibility coefficient, $K$,
i.e., derived as a stiffness coefficient with respect to a change in the
bulk density, and the dynamic one, $K_{\omega }^{\prime }$ associated with
the zero sound velocity, see Eq. (\ref{2.13}). The FSD effect is responsible
for the collisional relaxation of the collective modes in the Fermi liquid
and for the non-Markovian character of the nuclear matter viscosity (memory
effect in the viscosity $\eta _{\omega }$, Eq. (\ref{2.14})). The memory
effects in the viscosity play an essential role in the description of the
temperature dependence of the refraction coefficient $n,$ see Eqs. (\ref
{3.nkapa}) and (\ref{3.tau}). We have noted also the bell-shaped form of the
attenuation coefficient $\kappa $ as a function of the temperature $T,$ see
Figs. 1 and 2 . This peculiarity of $\kappa (T)$ provides a new criterion
for the determination of the transition temperature $T_{{\rm tr}}$ between
the zero-sound and first-sound regimes in hot nuclear matter.

Our consideration provides a good basis for understanding the difference in
the development of the spinodal instability in nuclear matter taking
into account both the FSD and the viscosity effect. We have shown that the
instability growth rate $\Gamma (q)$ in an unstable nuclear matter with
velocity dependent effective interparticle interaction is a non-monotonous
function of the wave number $q$ because of the anomalous dispersion term.
The anomalous dispersion term removes an unphysical infinite growth of the
short wave length fluctuations. The non-monotonous behavior of the
instability growth rate $\Gamma (q)$ is characterized by two 
wave numbers $q_{{\rm \max }}$ and $q_{{\rm crit}}$. We point out that both
the FSD effect and the relaxation processes lead to a shift of $q_{{\rm \max 
}}$ to the longer wave length region, providing an increase of the relative
yield of heavier clusters.

The main results were obtained assuming a quadrupole distortion of the Fermi
surface. We point out, however, that the key expression (\ref{2.14}) for the
viscosity coefficient can be rewritten identically in the following form 
\begin{equation}
\eta _{\omega }=\frac{3}{4}m\rho _{{\rm eq}}(c_{0}^{2}-c_{1}^{2})%
\mathop{\rm Re}%
\left( \frac{\tau _{r,\omega }}{1-i\omega \tau _{r,\omega }}\right) ,
\label{5.1}
\end{equation}
where $c_{1}$ and $c_{0}$\ are the first and zero sound velocity
respectively. This expression can also be established from a general
consideration, see Ref. \cite{lali2} Ch.8, and can be used for arbitrary
multipolarities of the Fermi surface distortion.

\section{Acknowledgements}

This work was supported in part by the US Department of Energy under grant
\# FG03-93ER40773. We are grateful for this financial support. One of us
(V.M.K.) thank the Cyclotron Institute at Texas A\&M University for the kind
hospitality.

\newpage

\begin{figure}[tbp]
\centerline{\epsffile{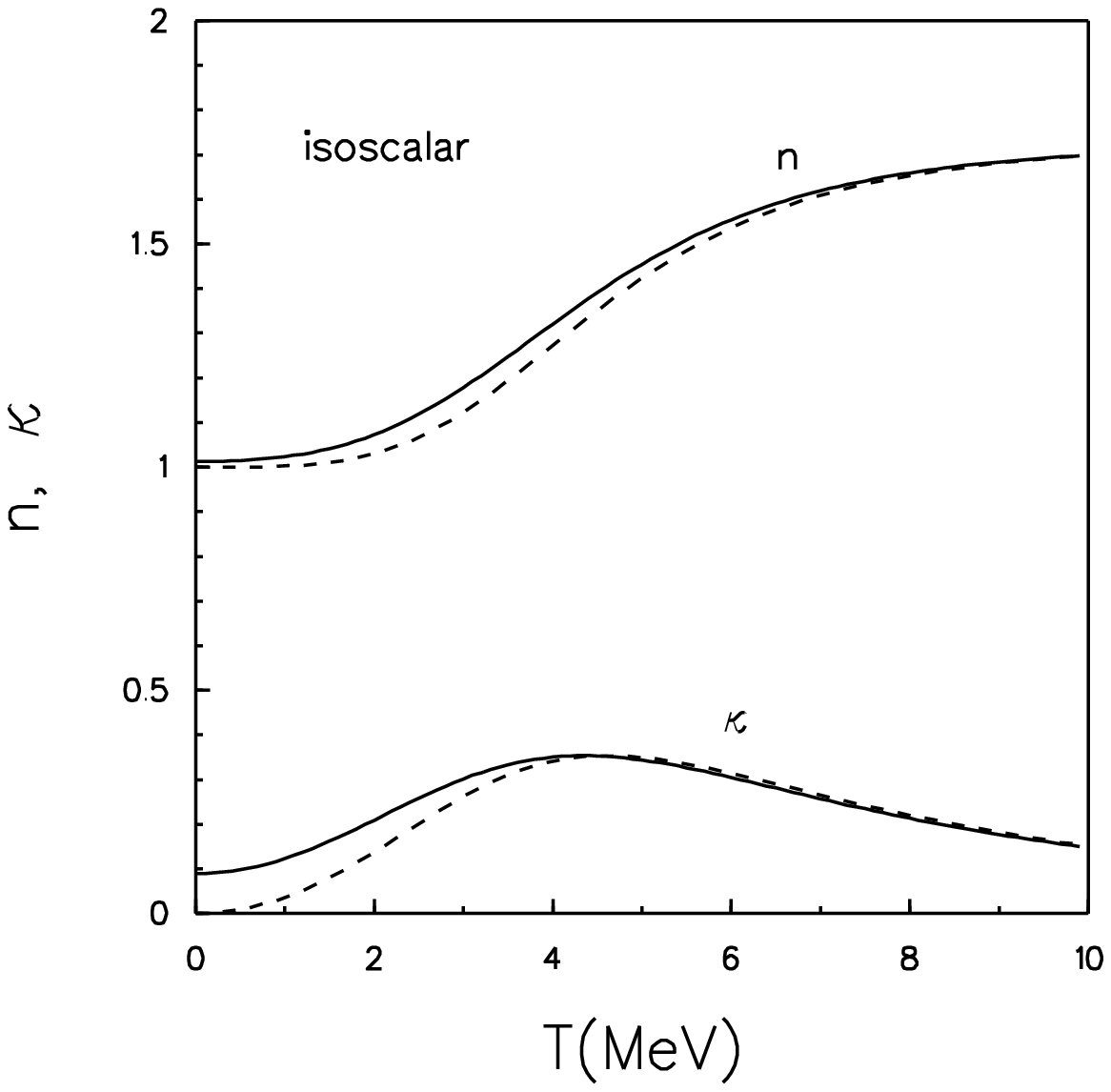}}
\caption{Refraction, $n$, and attenuation, $\protect\kappa$, coefficients of
the isoscalar sound wave as functions of temperature. The calculation was
performed for two eigenenergies $\hbar\protect\omega=1 {\rm MeV}$ (solid
line) and $\hbar\protect\omega=1 {\rm eV}$ (dashed line).}
\end{figure}

\begin{figure}[tbp]
\centerline{\epsffile{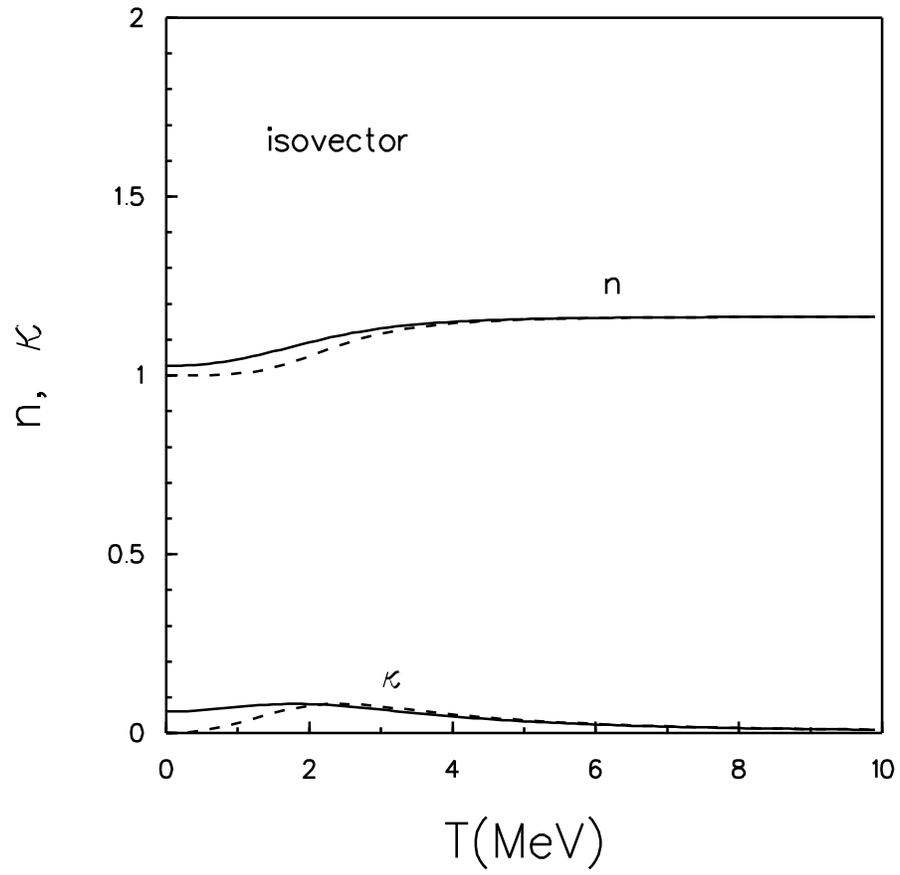}}
\caption{Same as Fig. 1 for isovector mode.\qquad\qquad\qquad\qquad\qquad}
\end{figure}

\begin{figure}[tbp]
\centerline{\epsffile{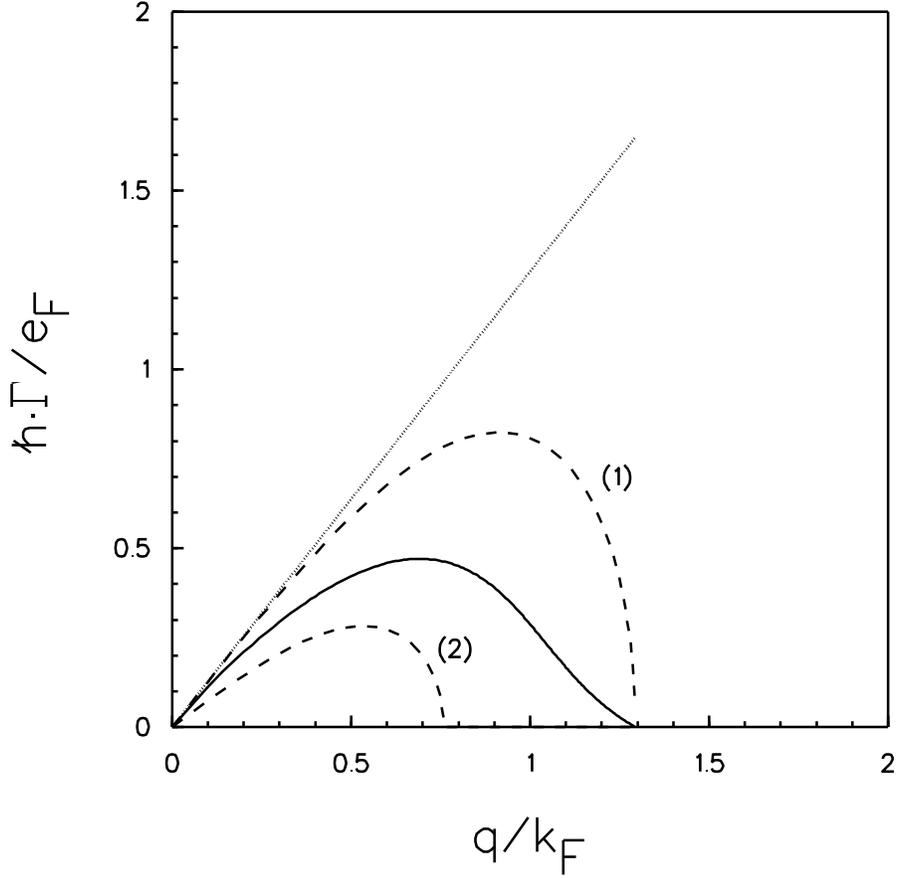}}
\caption{Dependence of the instability growth rate $\Gamma$ on the wave
number $q$. The calculations were performed for Skyrme force SIII,
temperature $T=6\,$MeV and density $\protect\rho_0 = x\,\protect\rho_{{\rm eq%
}}$ with $x=0.3$ and $\protect\rho_{{\rm eq}}=0.1453\,{\rm fm}^{-3}$. The
solid curve is for the viscous nuclear matter with $\protect\alpha = 9.2\,$%
MeV including both the memory and the Fermi-surface distortion effects. The
dashed and dotted lines are the results for the nonviscous liquid: curve (1)
is for a nuclear matter neglecting the FSD effect; curve (2) is the result
in the presence of the FSD effect and dotted line is the same as curve (1)
neglecting the anomalous dispersion term.}
\end{figure}

\begin{figure}[tbp]
\centerline{\epsffile{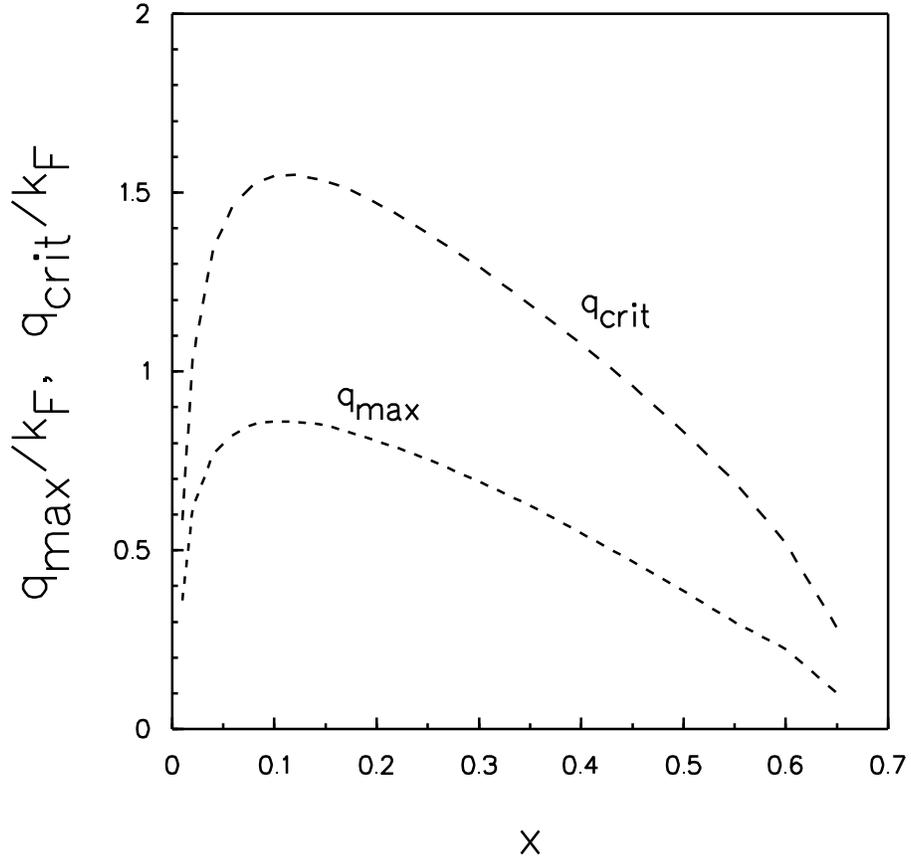}}
\caption{Dependence of the characteristic wave numbers of the instability 
growth
rate $\Gamma (q)$ on the dimensionless density parameter $x=\protect\rho_0/%
\protect\rho_{{\rm eq}}$. The calculation was performed for Skyrme force
SIII and temperature $T=6\,$MeV.}
\end{figure}

\end{document}